\documentclass[%
 reprint,
 amsmath,amssymb,
 aps,
]{revtex4-2}

\usepackage{graphicx}
\usepackage{dcolumn}
\usepackage{bm}
\usepackage{mathtools}
\usepackage{booktabs}

\usepackage{color}
\usepackage{multirow}
\usepackage[normalem]{ulem}
\usepackage[colorlinks=False,pdfusetitle]{hyperref}
\hypersetup{pdfauthor={Name}}

\newcommand{\MeV}{{\,\rm{MeV}}}

\definecolor{Green}{RGB}{0, 128, 0}
\graphicspath{{./Figures/}}

\begin{document}

\preprint{APS/123-QED}

\title{Electroweak Constraints on Nuclear Neutron Distributions and the Low-Energy Weak Mixing Angle}

\author{Xu-Run Huang$^{1,2}$}
\email{xr\_huang@cqu.edu.cn}

\author{Tong-Gang Yue$^{3}$}
\email{1076788852@sjtu.edu.cn}

\author{Zhen Zhang$^{4}$}
\email{zhangzh275@mail.sysu.edu.cn}

\author{Lie-Wen Chen$^{3}$}
\email{lwchen@sjtu.edu.cn}

\affiliation{$^{1}$Department of Physics, Chongqing Key Laboratory for Strongly Coupled Physics,
Chongqing University, Chongqing 401331, China
\\
$^{2}$Shanghai Research Center for Theoretical Nuclear Physics, NSFC and Fudan University, Shanghai 200438, China
\\
$^{3}$State Key Laboratory of Dark Matter Physics, Key Laboratory for Particle Astrophysics and Cosmology (MOE),
and Shanghai Key Laboratory for Particle Physics and Cosmology,
School of Physics and Astronomy, Shanghai Jiao Tong University, Shanghai 200240, China
\\
$^{4}$Sino-French Institute of Nuclear Engineering and Technology, Sun Yat-Sen University, Zhuhai 519082, China}

\date{\today}

\begin{abstract}
Coherent elastic neutrino--nucleus scattering (CE$\nu$NS) offers a low-energy probe of the weak mixing angle, $\sin^2\theta_W$, and possible new physics in the neutrino sector.
Its reliable interpretation, however, requires independent constraints on the neutron distributions of the target nuclei.
We exploit correlations across nuclei within an extended Skyrme energy-density-functional framework to propagate constraints from the CREX and PREX-II parity-violating electron-scattering (PVES) measurements to CE$\nu$NS targets.
This yields neutron form factors and radii, with relative radius uncertainties below $1\%$.
Using these inferred neutron form factors, we jointly analyze the COHERENT CsI, Ge, and Ar data and obtain $\sin^2\theta_W=0.231^{+0.016}_{-0.015}$, consistent with the Standard Model prediction.
These results provide the first electroweak-experimentally constrained benchmarks for nuclear-structure studies, neutrino-sector electroweak tests, and CE$\nu$NS signal predictions in dark-matter detectors.
\end{abstract}

\maketitle


\textit{Introduction.}---
Low-energy precision measurements, including neutrino scattering measurements and direct dark-matter (DM) searches, provide a complementary frontier for testing the Standard Model (SM) and searching for weakly coupled new physics~\cite{Dutta:2019oaj,Billard:2021uyg,Pandey:2023arh}.
Coherent elastic neutrino-nucleus scattering (CE$\nu$NS) is a neutral-current process whose cross section scales approximately as the square of the neutron number through coherent enhancement~\cite{Freedman:1973yd}.
Following its first observation by the COHERENT Collaboration using CsI at the Spallation Neutron Source (SNS)~\cite{COHERENT:2017ipa}, further CE$\nu$NS measurements have been reported with liquid-argon, CsI, and germanium targets~\cite{COHERENT:2020iec,COHERENT:2021xmm,COHERENT:2026yje,COHERENT:2024axu}, as well as with reactor antineutrinos~\cite{Ackermann:2025obx} and with solar $^{8}\mathrm{B}$ neutrinos in DM direct-detection experiments~\cite{PandaX:2024muv,XENON:2024ijk,XENON:2026smt}.
These measurements offer access to the weak mixing angle, $\sin^2\theta_W$, a central electroweak parameter that governs neutral-current couplings~\cite{Huang:2019ene,DeRomeri:2022twg,Maity:2024aji,Huang:2024jbh,Yip:2025dxe,DeRomeri:2026dac,AtzoriCorona:2026wbu}.
Its low-energy determination tests the SM prediction for its running~\cite{Kumar:2013yoa,ParticleDataGroup:2024cfk}, while a discrepancy could point to new physics in the neutrino sector~\cite{Abdullah:2022zue,Barranco:2005yy,Lindner:2016wff,Huang:2019ene,Billard:2018jnl,Cadeddu:2018dux,Liao:2024qoe,Breso-Pla:2023tnz,Dutta:2019eml,AtzoriCorona:2025xwr}.

The CE$\nu$NS cross section depends on both $\sin^2\theta_W$ and the nuclear weak response, leading to correlated constraints on the weak mixing angle and the neutron distribution in the target nuclei~\cite{Huang:2019ene}.
Fixing $\sin^2\theta_W$ to its SM value allows CE$\nu$NS data to constrain the point-neutron rms radius $R_n$~\cite{Cadeddu:2017etk,Coloma:2020nhf,Huang:2022wqu,Sierra:2023pnf}.
Although the point-proton rms radius $R_p$ is tightly constrained by electromagnetic measurements~\cite{DeVries:1987atn,Fricke:1995zz,Angeli:2013epw}, $R_n$ remains less certain because hadronic extractions carry considerable systematic uncertainties~\cite{Thiel:2019tkm,Mammei:2023kdf}.
The associated neutron skin thickness, $R_n-R_p$, is a fundamental property of nuclear structure, linking finite nuclei to the symmetry energy of neutron-rich matter and the structure of neutron stars~\cite{Brown:2000pd,Typel:2001lcw,Horowitz:2000xj,Furnstahl:2001un,Chen:2005ti,Chen:2010qx,Centelles:2008vu,Roca-Maza:2011qcr,Zhang:2013wna,Reinhard:2021utv,Hu:2021trw}.
It has been established that parity-violating electron scattering (PVES) provides a clean electroweak probe of neutron distributions through the nuclear weak form factor $F_W^e$~\cite{Thiel:2019tkm,Horowitz:1999fk}.
The PREX-II and CREX PVES measurements have constrained the charge--weak form-factor difference, $\Delta F_{\mathrm{CW}}=F_{\mathrm{ch}}-F_W^e$, in $^{208}\mathrm{Pb}$ and $^{48}\mathrm{Ca}$, respectively, enabling extractions of $R_n$ with $\sim\!1\%$ precision~\cite{Abrahamyan:2012gp,PREX:2021umo,CREX:2022kgg}.
However, describing both results within a common nuclear energy-density-functional (EDF) framework remains challenging, giving rise to the CREX--PREX puzzle~\cite{CREX:2022kgg,Reinhard:2022inh,Yuksel:2022umn,Zhang:2022bni,Papakonstantinou:2022gkt, Mondal:2022cva,Miyatsu:2023lki}.
Furthermore, since the PREX-II and CREX probe only two nuclei at specific momentum transfers, another challenge is to propagate their constraints to the neutron form factors of CE$\nu$NS targets over the relevant momentum-transfer range~\cite{Mammei:2023kdf,Piekarewicz:2025lel}.

Recent EDF studies have shown that an enhanced isovector spin--orbit (IVSO) interaction can reconcile the CREX and PREX-II constraints~\cite{Yue:2024srj,Qiu:2025qbv,Zhao:2024gjz}.
Building on this insight, we construct an extended Skyrme EDF family in this work, that incorporates this enhanced IVSO interaction with different density dependence of symmetry energy~\cite{Yue:2021yfx,Yue:2024srj}.
Within this family, correlations in the charge--weak form-factor difference across nuclei allow us to infer neutron form factors for CE$\nu$NS targets from the CREX and PREX-II measurements, yielding rms radii with relative uncertainties below $1\%$. The resulting neutron-distribution constraints break the CE$\nu$NS-only degeneracy between $\sin^2\theta_W$ and the neutron response.
Using these inferred neutron form factors, for the first time, we jointly analyze the COHERENT CsI, Ge, and Ar data~\cite{COHERENT:2021xmm,COHERENT:2026yje,COHERENT:2020iec} and obtain a low-energy determination of $\sin^2\theta_W$ with a precision of $6.8\%$, consistent with the SM prediction.
This precision gain is driven primarily by the latest Ge data, while the PVES constraints place the nuclear uncertainty estimates on a firmer experimental footing.

\textit{Nuclear weak response.}---
The nuclear weak response probed by CE$\nu$NS and PVES depends on both nuclear structure and electroweak couplings.
For a nucleus with $Z$ protons and $N$ neutrons, the nuclear weak charge is $Q_W^\xi=ZQ_W^{p,\xi}+NQ_W^{n,\xi}$, with $\xi=e$ for PVES and $\xi=\nu_\ell$ for CE$\nu$NS.
The nucleon weak charges are related to their vector couplings by $Q_W^{t,\xi}=2g_V^{t,\xi}$, with $t=p,n$.
At tree level, $g_V^p=1/2-2\sin^2\theta_W$ and $g_V^n=-1/2$, so the nuclear weak charge depends on $\sin^2\theta_W$ through the proton coupling.
We include electroweak radiative corrections in the $\overline{\mathrm{MS}}$ scheme, which introduce process dependence and, for neutrino scattering, flavor dependence~\cite{Erler:2013xha}.

The nuclear weak form factor $F_W^\xi$ at three-momentum transfer $q$ is given by~\cite{Reinhard:2021utv,Yue:2024srj}
\begin{align}\label{eq:Fw_GF0}
Q_W^\xi F_W^\xi(q)
={}& Z\,\widetilde{G}_E^{p,\xi}(q)F_p(q) +N\,\widetilde{G}_E^{n,\xi}(q)F_n(q) \notag\\
&+\sum_{t=p,n} \widetilde{G}_M^{t,\xi}(q)F_t^{(ls)}(q),
\end{align}
where $F_{p(n)}$ is the point-proton (neutron) form factor, $\widetilde{G}_{E(M)}^{t,\xi}$ is the intrinsic weak electric (magnetic) form factors of the nucleons, and $F_t^{(ls)}$ denotes the spin--orbit current response.
The form factors $F_p$, $F_n$, and $F_W^\xi$ are normalized to unity at $q=0$.
It should be noted that, for a given nucleus, CE$\nu$NS and PVES probe the same point-nucleon distributions but involve process-dependent nuclear weak form factors.

In CE$\nu$NS, a neutrino scatters coherently from the nucleus as a whole at low $q$.
For a nucleus $A$ of mass $M$, the SM differential cross section for a neutrino of flavor $\ell$ can be written as~\cite{Lindner:2016wff,Hoferichter:2020osn}
\begin{equation}
\frac{\mathrm{d}\sigma_{\nu_\ell A}}{\mathrm{d}T} = 
\frac{G_F^2 M}{4\pi}\left(1-\frac{T}{E_{\nu}}-\frac{MT}{2E_\nu^2}\right)
\left|Q_W^{\nu_\ell}F_W^{\nu_\ell}(q)\right|^2,
\end{equation}
where $G_F$ is the Fermi constant and $E_\nu$ is the incident neutrino energy.
The nuclear recoil energy $T$ satisfies $0\leq T\leq 2E_\nu^2/(M+2E_\nu)$, with $q^2 \simeq 2MT$.

For the typical momentum transfers in stopped-pion CE$\nu$NS experiments, the spin--orbit terms in Eq.~\eqref{eq:Fw_GF0} have a negligible impact on the present analysis~\cite{Hoferichter:2020osn}\footnote{We note that including these terms increases the total predicted CE$\nu$NS yield by $\sim\!0.4\%$ for CsI and $\sim\!0.2\%$ for Ar and Ge, with no appreciable change in the extracted $\sin^2\theta_W$.}.
We therefore neglect these terms in the CE$\nu$NS analysis but retain them in the EDF calculations of $F_W^e$.
Expanding $\widetilde{G}_E^{t,\nu_\ell}(q)$ to order $q^2$ gives
\begin{equation}\label{eq:QwFw_f}
\begin{aligned}
Q_W^{\nu_\ell}F_W^{\nu_\ell}(q) \simeq{}& ZQ_W^{p,\nu_\ell}
\left(1-\frac{q^2}{6} \langle\tilde r_p^2\rangle_{\nu_\ell}\right)F_p(q)
\\
&+NQ_W^{n,\nu_\ell} \left(1-\frac{q^2}{6}\langle\tilde r_n^2\rangle_{\nu_\ell}\right)F_n(q),
\end{aligned}
\end{equation}
where $\langle\tilde r_t^2\rangle_{\nu_\ell}$ denotes the nucleon weak-charge mean-square radius, expressed in terms of the proton, neutron, and strange electric mean-square radii.
Explicit expressions for these weak radii and the radiative corrections to $Q_W^{t,\xi}$, together with the parameters, are provided in Sec.~S2 of the Supplemental Material (SupMat).

\textit{PVES-informed neutron form factors and radii.}---
The CREX and PREX-II PVES measurements determine the differences between the nuclear charge and weak form factors, $\Delta F_{\rm CW}=F_{\rm ch}-F_W^e$, for $^{48}\mathrm{Ca}$ and $^{208}\mathrm{Pb}$, respectively~\cite{CREX:2022kgg,PREX:2021umo}:
\begin{align}
{}^{48}\mathrm{Ca}:\quad \Delta F_{\rm CW} &= 0.0277(55), & q &= 172.3\,\mathrm{MeV}, \\
{}^{208}\mathrm{Pb}:\quad \Delta F_{\rm CW} &= 0.041(13), & q &= 78.48\,\mathrm{MeV}.
\end{align}
Using these measurements as constraints, we extract the neutron form factors of CE$\nu$NS targets through EDF-based correlations across nuclei.

\begin{figure}[t!]
 \centering
 \includegraphics[width=0.98\columnwidth]{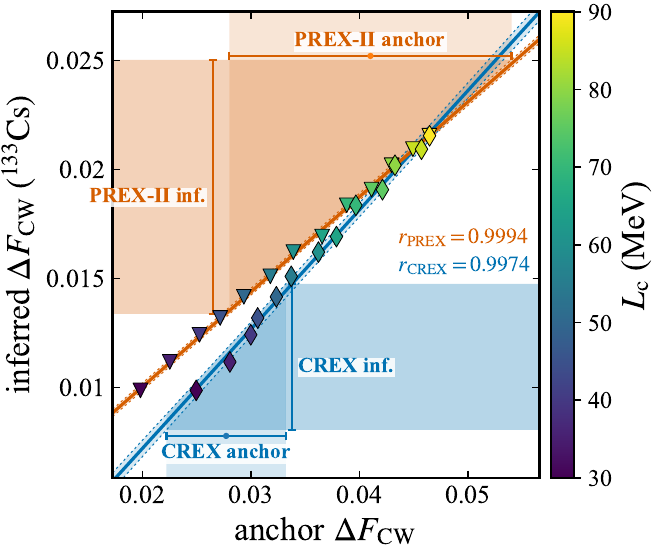}
 \caption{Illustration of inferring constraints on $\Delta F_{\rm CW}$ for $^{133}\mathrm{Cs}$ at $q = 50\,\mathrm{MeV}$. The correlation between $\Delta F_{\rm CW}$ for $^{133}\mathrm{Cs}$ and that for $^{48}\mathrm{Ca}$ (CREX) and $^{208}\mathrm{Pb}$ (PREX-II) is displayed with colored diamonds and triangles, respectively. Each point corresponds to a prediction from our EDFs with different symmetry-energy slope parameter $L_c$ at the subsaturation density $\rho_c=0.11\,\mathrm{fm}^{-3}$. The slanted bands, delimited by dashed lines, indicate the $1\sigma$ prediction uncertainties of the linear fits (solid lines); the Pearson correlation coefficients $r_{\text{CREX}}$ and $r_{\text{PREX}}$ are also shown. See text for details.
}
\label{fig:method}
\end{figure}

Physically, $\Delta F_{\rm CW}$ depends strongly on the symmetry-energy slope parameter $L_c$ at the subsaturation density $\rho_c=0.11\,\mathrm{fm}^{-3}$, since $L_c$ plays a key role in determining the neutron skin thickness of neutron-rich nuclei~\cite{Brown:2000pd,Typel:2001lcw,Zhang:2013wna,Qiu:2025qbv}.
This common dependence underlies strong correlations in $\Delta F_{\rm CW}$ across different nuclei.
To quantify these correlations, we construct an extended Skyrme EDF family (see SupMat Sec.~S1 for details) based on the framework of Ref.~\cite{Yue:2021yfx}, incorporating the enhanced isovector spin--orbit interaction suggested in Ref.~\cite{Yue:2024srj}.
The calibration uses finite-nucleus properties, nuclear-matter constraints from heavy-ion collisions, and the observed maximum neutron-star mass, with $L_c$ varied from 30 to $90\MeV$~\cite{Yue:2021yfx}.
The weak charges $Q_W^{t,e}$ are fixed to their SM values~\cite{Reinhard:2021utv}.
For $q$ below the first zero of $F_W^e$ for each target nucleus, the EDF predictions for $\Delta F_{\rm CW}$ exhibit strong linear correlations with those for $^{48}\mathrm{Ca}$ and $^{208}\mathrm{Pb}$ at their respective experimental momentum transfers.

As an example, figure~\ref{fig:method} illustrates these correlations and the propagation of the PVES constraints for $^{133}\mathrm{Cs}$ at $q=50\,\mathrm{MeV}$, with Pearson correlation coefficients $r_{\rm CREX},r_{\rm PREX}>0.99$.
For each target nucleus and momentum transfer $q$, we fit the EDF predictions for $\Delta F_{\rm CW}(q)$ as a linear function of those for $^{48}\mathrm{Ca}$ or $^{208}\mathrm{Pb}$ at the respective experimental momentum transfer.
These relations map the CREX and PREX-II measurements into two constraints on the target $\Delta F_{\rm CW}(q)$.
For each constraint, the $1\sigma$ uncertainty combines the propagated experimental uncertainty and the regression prediction error in quadrature~\cite{DraperSmith1998}.
Assuming independent Gaussian uncertainties, we combine the two constraints into a final PVES-informed band.
The corresponding weak form factor is then obtained as $F_W^e(q)=F_{\rm ch}(q)-\Delta F_{\rm CW}(q)$, with $F_{\rm ch}(q)$ determined from experimental charge-density information~\cite{DeVries:1987atn,Fricke:1995zz}.

To extract the point-neutron form factor from the PVES-informed $F_W^e(q)$, we calculate the ratio $\mathcal{R}_n^e(q)=F_n^{\rm EDF}(q)/F_W^{e,\rm EDF}(q)$ for each EDF parametrization and obtain $F_n(q)=\mathcal{R}_n^e(q)F_W^e(q)$.
The baseline ratio is taken from the IVSO250-Lc50 EDF, which reproduces both PVES measurements within their respective $1\sigma$ uncertainties.
The adopted $L_c = 50\MeV$ value represents a compromise between the stiffer symmetry energy favored by PREX-II and the softer symmetry energy favored by CREX, the electric dipole polarizabilities of $^{48}$Ca and $^{208}$Pb, and chiral effective field theory predictions for pure neutron matter~\cite{Yue:2024srj}.
As a consistency check, we also reconstruct the point-proton form factor as $F_p'(q)=\mathcal{R}_p^e(q)F_{\rm ch}(q)$, where $\mathcal{R}_p^e(q)=F_p^{\rm EDF}(q)/F_{\rm ch}^{\rm EDF}(q)$.
Across the IVSO250-Lc family, with $L_c$ ranging from 30 to 90 MeV, we note the central values of the neutron and proton rms radii extracted using these ratios differ by less than $0.2\%$ from those obtained with the baseline EDF.

We consider target materials relevant to CE$\nu$NS, PVES, and dark-matter experiments: argon~\cite{COHERENT:2020iec,DarkSide20k:2020ymr}, calcium~\cite{CREX:2022kgg}, germanium~\cite{Ackermann:2025obx,COHERENT:2026yje}, xenon~\cite{PandaX:2024muv,XENON:2024ijk}, CsI~\cite{COHERENT:2021xmm,COHERENT:2022nrm,An:2025lws}, and lead~\cite{RES-NOVA:2021gqp,PREX:2021umo,Becker:2018ggl}.
For each target nucleus, the PVES-informed $F_n$ is fitted with the Helm form factor~\cite{Helm:1956zz} to determine the neutron rms radius $R_n$ and surface width $s_n$ simultaneously.
Table~\ref{tab:helm_main} lists the fitted parameters for representative nuclei.
We note the relative uncertainties on the radii, which include the contribution propagated from the PVES-informed form-factor bands and the model uncertainty estimated by varying the key EDF parameters, remain below $1\%$ for all nuclei considered.
For the CE$\nu$NS analysis, the largest uncertainty within each target material is conservatively assigned to all its constituent isotopes.
The complete results and fitting details are provided in SupMat Sec.~S3.

\begin{table}[t]
\caption{
Helm parameters $(R_n,s_n)$ from fits to the central PVES-informed neutron form-factor curves for representative nuclei.
The relative uncertainty in $R_n$ is $\sigma^2_n=\sigma^2_b+\sigma^2_m$,
where $\sigma_{b}$ is the propagated neutron-radius band half-width and $\sigma_m = 0.2\%$ is the estimated model uncertainty.
}
\label{tab:helm_main}
\begin{ruledtabular}
\begin{tabular}{lccc}
Isotope & $R_n$ [fm] & $s_n$ [fm] & $\sigma_{n}$ [\%] \\
\hline
$^{40}\mathrm{Ar}$  & 3.417 & 0.970 & 0.46 \\
$^{74}\mathrm{Ge}$  & 4.118 & 1.057 & 0.60 \\
$^{127}\mathrm{I}$  & 4.822 & 1.009 & 0.67 \\
$^{132}\mathrm{Xe}$ & 4.879 & 1.029 & 0.72 \\
$^{133}\mathrm{Cs}$ & 4.880 & 1.012 & 0.68 \\
$^{208}\mathrm{Pb}$ & 5.639 & 1.053 & 0.79 \\
\end{tabular}
\end{ruledtabular}
\end{table}

An analogous Helm fit to $F_p'$ yields the point-proton rms radius $R_p'$ and surface width $s_p'$.
The radius $R_p'$ is compared with the reference radius $R_p$ inferred from the measured nuclear charge radius $R_{\rm ch}$~\cite{Angeli:2013epw}.
The relative difference $|R_p'-R_p|/R_p$ averages $0.09\%$ and remains below $0.34\%$, supporting the consistency of the EDF-based scaling method.

\begin{figure*}[t!]
 \centering
 \includegraphics[width=0.8\textwidth]{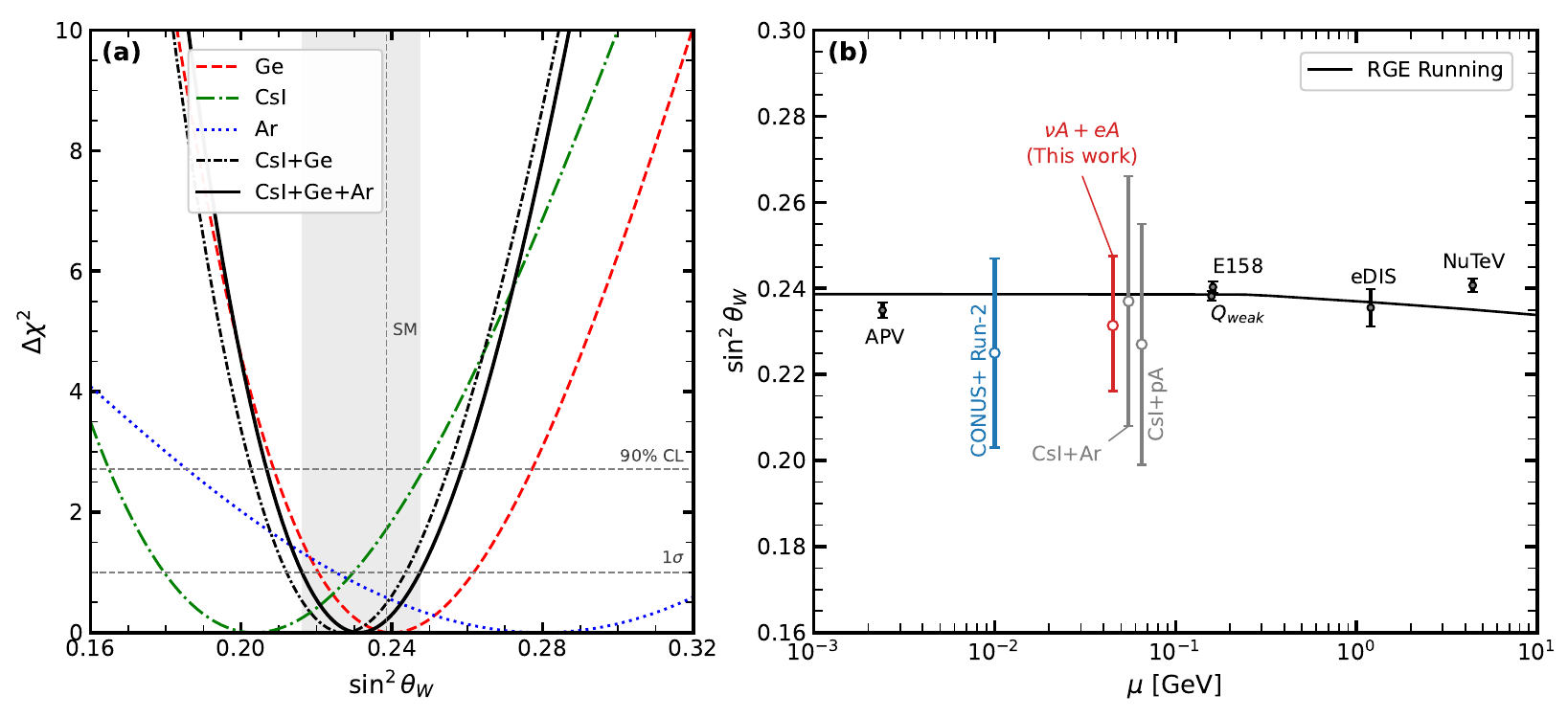}
 \caption{\label{fig:results}
    PVES-informed determination of the low-energy $\sin^2\theta_W$ from COHERENT data.
    (a) $\Delta\chi^2\equiv\chi^2-\chi^2_{\min}$ profiles of $\sin^2\theta_W$ from fits to the COHERENT Ge, CsI, and Ar data~\cite{COHERENT:2021xmm,COHERENT:2020iec,COHERENT:2026yje}. Colored curves show the individual fits and the black curves their joint fits. The horizontal dashed lines indicate $1\sigma$ and $90\%$ C.L., and the vertical dashed line marks the SM value.
    (b) $\sin^2\theta_W$ determinations at representative momentum-transfer scales $\mu$. The red open circle ($\nu A+eA$) shows our $1\sigma$ result. 
    Blue and gray open circles denote previous CE$\nu$NS-based extractions of $\sin^2\theta_W$ from CONUS+~\cite{Ackermann:2026mms}, CsI+$pA$~\cite{Huang:2024jbh}, and CsI+Ar~\cite{DeRomeri:2022twg}.
    SNS-based points are slightly offset horizontally for clarity. 
    Filled black points show other experimental results~\cite{Qweak:2018tjf,NuTeV:2001whx,PVDIS:2014cmd,SLACE158:2005uay,Dzuba:2012kx}, and the solid curve is the SM $\overline{\mathrm{MS}}$ running~\cite{ParticleDataGroup:2024cfk}.
    }
\end{figure*}
\textit{Low-energy $\sin^2\theta_W$ from COHERENT data.}---
Although recent measurements with reactor antineutrinos and solar $^{8}\mathrm{B}$ neutrinos have broadened the CE$\nu$NS program~\cite{Ackermann:2025obx,PandaX:2024muv,XENON:2024ijk,XENON:2026smt}, the full COHERENT CsI and latest Ge measurements remain the only currently available CE$\nu$NS results with flux-averaged cross sections determined at the $\sim\!15\%$ level~\cite{COHERENT:2021xmm,COHERENT:2026yje}.
We also include the less precise CENNS-10 Ar measurement~\cite{COHERENT:2020iec}.
All three measurements use the pulsed pion-decay-at-rest neutrino source at the SNS, which produces a prompt $\nu_\mu$ flux and delayed $\nu_e$ and $\bar{\nu}_\mu$ fluxes.
The pulsed time structure helps distinguish the prompt and delayed contributions while strongly suppressing steady-state backgrounds.

Using the fitted Helm form factors, we determine $\sin^2\theta_W$ by fitting the reconstructed-energy and arrival-time distributions for each measurement and for the CsI+Ge and CsI+Ge+Ar combinations.
Experimental and neutron-radius uncertainties are included (see SupMat Sec.~S4 for details).
Figure~\ref{fig:results}(a) shows the $\Delta\chi^2=\chi^2-\chi^2_{\min}$ profiles, where $\chi^2_{\min}$ is the minimum for each fit.
One sees the $1\sigma$ intervals of the individual fits overlap.
The CsI and Ge data provide the core constraint, with their joint fit yielding $\sin^2\theta_W=0.227^{+0.016}_{-0.015}$.
The broader Ar profile favors higher values, and its inclusion shifts the combined best fit upward while leaving the $1\sigma$ width nearly unchanged.
The three-target fit yields
\begin{equation}
\sin^2\theta_W=0.231^{+0.016}_{-0.015}.
\end{equation}
The SM prediction has $\Delta\chi^2=0.2$ in the combined profile and lies well within the $1\sigma$ interval.

Figure~\ref{fig:results}(b) places our PVES-informed CE$\nu$NS determination (the $\nu A+eA$ point) alongside the SM running of $\sin^2\theta_W$ and existing atomic, electron-scattering, and neutrino-scattering results~\cite{Kumar:2013yoa,ParticleDataGroup:2024cfk}.
The gray points show two earlier COHERENT-based CE$\nu$NS extractions.
The CsI+Ar extraction is based on a joint fit using a Klein--Nystrand form factor with an empirical nuclear-radius prescription~\cite{DeRomeri:2022twg}, whereas the CsI+pA extraction adopts a neutron radius inferred from proton--nucleus elastic scattering, a hadronic probe~\cite{Huang:2024jbh}.
In contrast, our EDF-based analysis constrains the target neutron distributions using the CREX and PREX-II PVES measurements.
Our $1\sigma$ uncertainty on $\sin^2\theta_W$ is nearly a factor of two smaller than those of the two earlier extractions, primarily owing to the inclusion of the latest Ge data.
At the momentum-transfer scale $\mu\!\sim\!10\,\mathrm{MeV}$, the CONUS Collaboration recently obtained $\sin^2\theta_W=0.225 \pm 0.022$ from reactor-antineutrino CE$\nu$NS, consistent with the SM prediction~\cite{Ackermann:2026mms}.
These neutrino-sector determinations remain less precise than the low-energy results from APV, $Q_{\rm weak}$, and SLAC E158 in the electron sector~\cite{Dzuba:2012kx,Qweak:2018tjf,SLACE158:2005uay}.
CE$\nu$NS nevertheless offers a distinct low-energy test of neutrino neutral-current interactions~\cite{Abdullah:2022zue}.
Our present results provide the first electroweak-experimentally constrained benchmark for the determination of the low-energy weak mixing angle, and a higher precision is anticipated with enhanced statistics from the ongoing CE$\nu$NS and PVES experiments.

\textit{Summary.}---
Uncertainties in the neutron distributions of target nuclei complicate the electroweak interpretation of CE$\nu$NS measurements.
Within an extended Skyrme EDF framework, we have inferred neutron form factors for CE$\nu$NS targets from the CREX and PREX-II PVES measurements.
The extracted neutron rms radii have relative uncertainties below $1\%$.
With these neutron form factors, a joint fit to the COHERENT CsI, Ge, and Ar data yields $\sin^2\theta_W=0.231^{+0.016}_{-0.015}$ at a representative momentum-transfer scale of $\mu\!\sim\!50\,\mathrm{MeV}$, consistent with the SM prediction.
Our work paves the way for improved low-energy determinations of $\sin^2\theta_W$ with future high-precision CE$\nu$NS and PVES experiments, including CICENNS~\cite{An:2025lws} and MREX~\cite{Becker:2018ggl}.
The inferred nuclear neutron distributions also provide the first electroweak-experimentally constrained benchmarks for nuclear-structure studies and CE$\nu$NS signal predictions in DM detectors.

\begin{acknowledgments}

The authors would like to thank Professor Paul-Gerhard Reinhard and Professor Gianluca Colo for helpful communications.
This work was supported in part by the National Natural Science Foundation of China under Grant Nos. 12235010, 12575137, and 12147101, the National SKA Program of China No. 2020SKA0120300, and the Science and Technology
Commission of Shanghai Municipality under Grant No. 23JC1402700.
\end{acknowledgments}


\bibliography{references}

\clearpage
\onecolumngrid  







\end{document}